\documentclass[11pt,twoside]{article}
\usepackage{graphicx}

\begin{document}

\bibliographystyle{IEEEtran} 

\title{A gravitational theoretical development supporting MOND}        


\author{Edmund A. Chadwick, Timothy F. Hodgkinson and Graham S. McDonald \\ 
{\it School of Computing, Science and Engineering },
{\it University of Salford} \\
{\it Salford M5 4WT, UK}
}


\date{\small{E.A.C. and T.F.H. contributed equally to the formulation, analyses and
interpretation. The authors are thus listed in alphabetic order. G.S.M.'s contribution was help with the point-source solutions.}}

\maketitle

\begin{abstract}
Conformal geometry is considered within a general relativistic framework.  An invariant distant for proper time is defined and a parallel displacement is applied in the distorted space-time, modifying Einstein's equation appropriately.  A particular solution is introduced for the covariant acceleration potential that matches the observed velocity distribution  at large distances from the galactic centre, i.e. Modified Newtonian Dynamics (MOND). This explicit solution, of a general framework that allows both curvature \textit{and} explicit local expansion of space-time, thus reproduces the observed flattening of galaxys' rotation curves without the need to assume the existence of dark matter. The large distance expansion rate is found to match the speed of a spherical shock wave. 
\end{abstract}

\section{Introduction}
The motion of stars around spiral galaxies trace flat rotation curves
which do not equate to those calculated from Newtonian dynamics
applied to the luminous matter of the galaxy \cite{Zwicky:1933}.
One possible explanation for this is the existence of non-luminous
dark matter such as Weakly Interacting Massive Particles \cite{LHCb:2012} which, 
when included in the calculation, 
reproduce the observed velocity profiles.
Observations made on the bullet cluster of galaxies suggest the
existence of dark matter \cite{Markevitzetal:2003},
although other effects have also been attributed to these observations
\cite{Milgrom:2012}.   
Another possible explanation is that galactic motions are governed by
non-Newtonian physics.
This viewpoint is backed by the Tully-Fisher relation
\cite{TullyFisher:1977} that shows a correlation between the speed of
rotation of stars and luminosity in a galaxy without the requirement
for dark matter.
One suggestion is a modification to Newtonian dynamics named MOND
\cite{Milgrom:1983} which produces the motions of spiral galaxies, 
and McGaugh \cite{McGaugh:2011} has demonstrated that it also fits the
motions for gas-rich galaxies.
A relativistic gravitation theory to support MOND dynamics has been
developed \cite{Bekenstein:2004},
although it has been suggested that this might lead to unstable
dynamics for stars \cite{Seifert:2007}.
Other theories have also been suggested, for example
conformal gravity \cite{Mannheim:1989} \cite{Mannheim:2006},
expanding space-time \cite{Masreliez:2012},
a theory based on curvature effects \cite{Capozziello:2006},
and a modification to the gravitational field equations
\cite{Ma:2012}. 
In the present paper, conformal geometry is used within a general
relativistic framework.
This formulation has similarities with Weyl theory \cite{Weyl:1918}
\cite{Einstein:1952} \cite{Eddington:1924} which considers a gauge
re-scaling that changes the vector length, 
and Weyl relates this to the electromagnetic potential satisfying
Maxwell's equations.  
However, this leads to a variance in the atomic time of clocks which
is not observed and
which led to the theory being discounted in particular by Einstein
\cite{Schulman:1997}.   
To overcome this, conformal gravity considers a variational in which
an infinitesimal gauge re-scaling occurs simultaneously with a conformal
transform that allows a counterbalancing length re-scaling such that
the line element remains invariant \cite{Mannheim:1989}
\cite{Mannheim:2006}. 
Similarly, the formulation presented here can equivalently be viewed
as a gauge re-scaling together with a length re-scaling to ensure that
for weak distortion of space-time the invariant (line element) proper
time is the atomic time and so does not vary as it does in Weyl theory.
Defining an invariant distance for proper time and applying parallel
displacement in the distorted space-time leads to a formulation that fits
MOND for the dynamics of galaxies by introducing a particular solution for the covariant
acceleration potential. 

\section{The distortion of space-time}

The notation and arguments as laid out by Dirac \cite{Dirac:1975} are
followed.
Assume that there exists a higher $N$-dimensional space described
by rectilinear contravariant coordinate points $z'^n (n=1,2,....N)$,
such that there is a distance measure $ds'$ between two neighbouring
points given by

\begin{equation}
ds'^2 = dz'_n dz'^n = h_{nm}dz'^m dz'^n,
\label{higherdimdist}
\end{equation}
where $dz'_n$ and $dz'^n$ are the covariant and contravariant
infinitesimal changes
in position respectively, and the tensor $h_{nm}$ is constant.
In the presence of matter, assume that this space is distorted by both
local expansions and curvature.

Consider local expansions first.
Allowing explicit expansions that are isotropic at point $z^m$ such that
$dz'^m=\sqrt{\alpha} dz^m$,
then (\ref{higherdimdist}) becomes
\begin{equation}
ds'^2 = \alpha h_{nm}dz^m dz^n,
\label{stretchdimdist}
\end{equation}
where the factor $1/\alpha $ is a function of position.

Now consider curvature.
In particular, consider a lower-dimensional curved `surface' lying in
the higher dimensional plane.
The lower-dimensional 4-space $x^{\mu}$, $(\mu =0,1,2,3)$ is defined, 
where $x^0$ denotes the time coordinate, and
$(x^1,x^2,x^3)$ denote the spatial coordinates. 
Follow the convention where Greek symbols denote indices summed from 0
to 3, 
and Roman symbols denote indices summed from 1.  
Let the point $y^n(x)$ in the higher dimensional plane correspond to a
point $x^{\mu}$ in four-dimensional space-time.
Then from (\ref{stretchdimdist}) we get
\begin{eqnarray}
ds'^2  
&=& \alpha h_{nm} y^n,_{\mu} y^m,_{\nu}
dx^{\mu} dx^{\nu} \nonumber \\
&=& \alpha y^n,_{\mu} y_n,_{\nu}
dx^{\mu} dx^{\nu} \nonumber \\
&=& g_{\mu \nu} dx^{\mu} dx^{\nu} 
\label{variantdist}
\end{eqnarray}
where the comma denotes a differentiation, and the
convention for inner product is $a^{\mu}b_{\mu}=
a^0b_0-a^1b_1-a^2b_2-a^3b_3$.  
The metric is therefore defined as
\begin{equation}
g_{\mu \nu}= \alpha y^n,_{\mu}y_n,_{\nu}.
\label{metricdefinition}
\end{equation}
Hence, the components of the metric tensor are determined by both
local expansion and curvature, from the $\alpha $ factor and from the
$y^n,_{\mu}y_n,_{\nu}$ factor respectively.

\section{Invariant distance and parallel displacement}
Requiring an invariant distance means that

\begin{equation}
ds^2 = y^n,_{\mu} y_n,_{\nu} dx^{\mu} dx^{\nu} ,
\label{defnds}
\end{equation}
where $ds$ is the invariant infinitesimal distance (line element or
proper time) between two infinitesimally close points.
For weak curvature (\ref{defnds}) becomes $ds = dx^0$, so proper
time becomes atomic time without a scaling factor present;
in Weyl theory, the distance measure is chosen as the rescaled gauge,
and so for weak curvature the scaling factor is present, 
leading to Einstein's objection. 
Equating this with (\ref{variantdist}) gives 

\begin{equation}
\alpha ds^2 = ds'^2 = \alpha y^n,_{\mu} y_n,_{\nu} dx^{\mu} dx^{\nu} ,
\label{dsalpha}
\end{equation}
which is equivalent to a gauge rescaling together with a
counterbalancing length rescaling on the left-hand-side and
right-hand-side of (\ref{dsalpha}) respectively.

The change in vector length due to parallel displacement is
\begin{equation}
dA_{\nu} = (A^{\mu} \alpha y^n,_{\mu}y_n,_{\nu  \sigma}+ A_{\nu} (\ln \alpha),_{\sigma})dx^{\sigma}.
\label{covariantchange}
\end{equation}
Note that when $\alpha = 1$ there is no expansion and the standard
result for parallel displacement is recovered. 

\section{Expansion symbols and covariant differentiation}
This analysis works on the metric and its re-scaling, 
so produces the same formulations obtained by Weyl for gauge
invariance \cite{Weyl:1918} \cite{Einstein:1952} \cite{Eddington:1924},
and give rise to expansion symbols $E_{\mu  \nu \sigma}$ and 
Christoffel symbols $\Gamma _{\mu \nu \sigma}$. The standard Christoffel symbol is,
$$
\Gamma _{\mu \nu \sigma} = 1/2(g_{\mu \nu},_{\sigma}
                          -g_{\sigma \nu},_{\mu} 
                          +g_{\mu \sigma},_{\nu} ),
$$
and the expansion symbol given by
\begin{equation}
E_{\mu \nu \sigma} = 1/2(g_{\mu \nu} (\ln \alpha ),_{\sigma} 
                    -g_{\sigma \nu} (\ln \alpha ),_{\mu}
                    +g_{\mu \sigma} (\ln \alpha ),_{\nu})
\end{equation}
are introduced.
So, the change in the covariant vector $A_{\nu}$ given by
(\ref{covariantchange}) can be rewritten as 
\begin{eqnarray}
dA_{\nu}&=& A^{\mu}(\Gamma_{ \mu \nu \sigma}-E_{ \mu \nu \sigma}
+g_{\mu \nu} (\ln \alpha),_{\sigma})dx^{\sigma} \nonumber \\
&=& A^{\mu} \Gamma_{ * \mu \nu \sigma} dx^{\sigma},
\label{covchange}
\end{eqnarray}
where the Christoffel symbol has been modified to 
\begin{eqnarray}
\Gamma_{* \mu \nu \sigma} &=& \Gamma _{\mu \nu \sigma} -E_{\mu \nu  \sigma}
+g_{\mu \nu} (\ln \alpha),_{\sigma} \nonumber \\
&=& \Gamma _{\mu \nu \sigma} +E_{\nu \mu  \sigma}
\end{eqnarray}

The infinitesimal change in the covariant vector is now used to define
covariant differentiation.
Noting that $A_{\mu}(x)+\Gamma ^{\alpha}_{*\mu \nu}A_{\alpha}
dx^{\nu}$ is a parallel displaced tensor so is also a tensor (where 
$g_{\alpha \beta} \Gamma ^{\alpha}_{* \mu \nu} = \Gamma _{* \beta \mu  \nu}$), 
then define a modified covariant derivative 

\begin{equation}
A_{\mu ; \nu}= A_{\mu},_{\nu} - \Gamma ^{\alpha} _{* \mu \nu}
A_{\alpha},
\label{codiffdef}
\end{equation}
as opposed to the standard covariant derivative given by
$A_{\mu : \nu}= A_{\mu},_{\nu} - \Gamma ^{\alpha} _{\mu \nu} A_{\alpha}$.
So the modified curvature tensor is 
$$
R^{\beta}_{* \nu \rho \sigma} = \Gamma ^{\beta}_{* \nu \sigma ,\rho} 
-\Gamma ^{\beta}_{* \nu \rho ,\sigma} 
+\Gamma ^{\alpha}_{* \nu \sigma} \Gamma ^{\beta}_{* \alpha \rho} 
-\Gamma ^{\alpha}_{* \nu \rho} \Gamma ^{\beta}_{* \alpha \sigma}.
$$
For weak curvature, dropping quadratics, this becomes
\begin{eqnarray}
R_{*\mu \nu} &=& 
\Gamma ^{\alpha}_{* \mu \alpha ,\nu} 
-\Gamma ^{\alpha}_{* \mu \nu  ,\alpha}  
\nonumber \\
&=& g^{\alpha \beta} \left(
\Gamma _{* \beta \mu \alpha ,\nu} 
-\Gamma _{* \beta \mu \nu  ,\alpha}  
\right) .
\label{weakcurv}
\end{eqnarray}

\section{Change in vector length and contravariant change}
The change in the dot product of two vectors is
\begin{eqnarray}
d(A^{\nu}B_{\nu}) &=& d(g^{\mu \nu} A_{\mu}B_{\mu}) \nonumber \\
&=& g^{\mu \nu}A_{\mu} dB_{\nu} +g^{\mu \nu}B_{\nu} dA_{\mu} +
A_{\mu} B_{\nu} g^{\mu \nu},_{\sigma} dx^{\sigma} . \nonumber
\end{eqnarray}
Substituting in for the covariant change (\ref{covchange}),
and using the fact that 
$\Gamma _{\mu \nu \sigma}+\Gamma _{\nu \mu \sigma}=g_{\mu
  \nu},_{\sigma}$
and 
$E_{\mu \nu \sigma}+E_{\nu \mu \sigma}=g_{\mu \nu}(\ln
\alpha),_{\sigma} $,
gives 
\begin{eqnarray}
d(A^{\nu}B_{\nu}) &=& \left[
A^{\nu}B^{\mu} g_{\mu \nu},_{\sigma} -A^{\nu}B^{\mu} g_{\mu \nu} (\ln
\alpha ),_{\sigma} \right. \nonumber \\
&& \left. 
+2A^{\nu} B_{\nu} (\ln \alpha ),_{\sigma} 
+A_{\alpha}B_{\beta} g^{\alpha \beta} ,_{\sigma} 
\right] dx^{\sigma}. \nonumber
\end{eqnarray}
Noting that $A_{\alpha }B_{\beta} g^{\alpha \beta}, _{\sigma}= 
            -A^{\nu }B^{\mu} g_{\mu \nu}, _{\sigma}$,
then gives
\begin{eqnarray}
d(A^{\nu}B_{\nu}) &=& A^{\nu} B_{\nu} (\ln \alpha),_{\sigma}
dx^{\sigma} \nonumber \\
&=& A^{\nu} B_{\nu} d(\ln \alpha). 
\label{dotprodinc}
\end{eqnarray}
Therefore,
$d(\frac{A^{\nu}B_{\nu}}{\alpha }) =0$
and so letting $A^{\nu}=B^{\nu}$ gives a change in vector length 
$$
d((1/\alpha )(A^{\nu} A_{\nu})) = 0,
$$
so the length of a vector changes by the factor $1/\alpha$ from point
to point.
Letting $A^{\nu} = dx^{\nu}$, then 
$d((1/\alpha )(dx^{\nu} dx_{\nu})) = d(ds'^2/\alpha)=0$,
giving
$$
d(ds)=0,
$$and so $ds$ is an invariant distance as expected from (\ref{defnds})
for consistency.

From (\ref{dotprodinc}), 
$d(A_{\nu}B^{\nu})=A_{\nu}B^{\nu}d(\ln \alpha) 
= A_{\nu}dB^{\nu}+dA_{\nu}B^{\nu}$,
this gives
\begin{eqnarray}
A_{\nu}B^{\nu} &=& d(A_{\nu}B^{\nu}) - A_{\nu}B^{\mu}
\Gamma ^{\nu}_{* \mu \sigma} dx^{\sigma} \nonumber \\
&=& A_{\nu}B^{\nu}d(\ln \alpha)
-A_{\nu}B^{\mu}\Gamma ^{\nu}_{* \mu \sigma} dx^{\sigma} . \nonumber
\end{eqnarray}
This holds for any $A_{\nu}$, and so cancelling the repeated term gives 
\begin{equation}
dB^{\nu} = - B^{\mu} \Gamma ^{* \nu}_{\mu \sigma } dx^{\sigma},
\label{contradefn}
\end{equation}
where $\Gamma ^{* \nu}_{\mu \sigma} =\Gamma ^{\nu}_{\mu \sigma} 
-g^{\nu}_{\mu} (\ln \alpha),_{\sigma}$, and so
$$
\Gamma ^* _{\alpha \mu \sigma} =
\Gamma  _{\alpha \mu \sigma} -
E _{\alpha \mu \sigma} .
$$

\section{Geodesic acceleration}
Letting
$$
dx^{\sigma} = \frac{dx^{\sigma}}{ds} ds = V^{\sigma} ds,
$$
where $ds$ is the invariant distance, then from (\ref{contradefn}) the
contravariant velocity $V^{\mu}$ in weak distorted space is
\begin{eqnarray}
\frac{dV^{\mu}}{ds} &=& -\Gamma ^{* \mu} _{\nu \sigma} V^{\nu}
V^{\sigma} \nonumber \\
\frac{dV^m}{ds} &=& -\Gamma ^{* m} _{\nu \sigma} V^{\nu} V^{\sigma} \nonumber \\
&=& -\Gamma ^{* m} _{00} V^0 V^0 \nonumber \\
&=& -g^{mn} \Gamma ^* _{n00} V^0 V^0 .
\label{geodesic}
\end{eqnarray}
For a static gravitational field, $g_{\mu \nu},_0 = \alpha ,_0=0$ and
also $g_{n0}=0$. 
So, $\Gamma _{n00}= (-1/2)g_{00},_n$ and $E_{n00}=(-1/2)g_{00}(\ln
\alpha),_n$.
Furthermore, from (\ref{variantdist}) $\alpha dx^2 = g_{\mu \nu}
dx^{\mu} dx^{\nu}$, and so for a static field (such that
$g_{m0}=g_{0m}=0$)
and for velocities small compared with light such that quadratics
$V^mV^n$ can be dropped, then $\alpha = g_{00} V^0 V^0$.  
Substituting these results into (\ref{geodesic}) gives
\begin{eqnarray}
\frac{dV^m}{ds} = -g^{mn} \Gamma ^*_{n00} V^0 V^0 &=&
(1/2)g^{mn}\left(
g_{00},_n V^0 V^0 - g_{00}(\ln \alpha),_n V^0 V^0 \right) \nonumber \\
&=& (1/2) (\alpha g^{mn}) \left( \ln (g_{00}/\alpha) \right),_n
\nonumber \\
&=& (\alpha g^{mn}) \phi ,_n , \nonumber
\end{eqnarray}
where 
\begin{equation}
\phi = \ln \sqrt{\frac{g_{00}}{\alpha}} 
\approx \sqrt{\frac{g_{00}}{\alpha}} - 1 
\approx (1/2) \left( \frac{g_{00}}{\alpha} - 1 \right)
\label{phidefn}
\end{equation}
is the covariant acceleration, since in the weak distortion limit
$\alpha g^{mn} = -1$ for $m=n$. 
It is noted that when $\alpha =1$, the standard result for geodesic
acceleration is obtained.

\section{Einstein's field equations and the gravitational force}
In empty space, Einstein's field equations then become
$$
R_{* \mu \nu} - (1/2)g_{\mu \nu} R_*= 0.
$$
In the presence of matter, a material energy tensor $T^{\mu \nu}$ is
required such that $T^{\mu \nu}_{; \mu}=0$, for the modified covariant
differentiation given by (\ref{codiffdef}).
Defining a velocity $V^{\mu}_*$ given by differentiating distance with
respect to the higher dimensional distance measure $s'$, 
then $V^{\mu }_* = dx^{\mu}/ds' = (1/\sqrt{\alpha }) V^{\mu}$ and so
$g_{\mu \nu} V^{\mu }_* V^{\nu}_* = 1$, leading to $V_{* \mu}V^{\nu
}_{* ; \sigma} = 0$.
Together with the condition for conservation of matter $(\rho V^{\mu
}_*)_{ ; \mu }$, then gives 
$T^{\mu \sigma }_{; \mu} = (\rho V^{\mu }_* V^{\nu } _*)_{; \mu}=0 $.
So, consider generalising Einstein's field in the presence of matter by
\begin{equation}
R_{* \mu \nu} - (1/2)g_{\mu \nu} R_*= -8\pi \rho V_{* \mu}V_{* \nu}.
\label{Einstein}
\end{equation}
For $\alpha =1$, $R_* = R$ and $V_{* \mu} = V_{\mu}$, and the standard
law is recovered. 
Rearranging in the usual way to incorporate the term $R_*$ into the
right hand side of (\ref{Einstein}), 
substituting for $R_{* \mu \nu} $ given by (\ref{weakcurv})
and neglecting quadratic quantities in $\Gamma $ and $E$ for weak
distortion, gives when $\mu=\nu=0$
$$
\alpha g^{\alpha \beta} 
\left( \Gamma _{\beta 0 \alpha}, _0 - \Gamma _{\beta 0 0}, _{\alpha}
\right)
+
\alpha g^{\alpha \beta} 
\left( E_{0 \beta \alpha}, _0 - E _{0 \beta 0 }, _{\alpha}
\right)
= -4\pi \rho  V_0 V_0 .
$$
A static field such that $g_{\alpha \beta},_0= (\ln \alpha ),_0=0$ 
gives $\Gamma _{\beta 0 \alpha },_0=0$,
$\Gamma _{\beta 0 0},_{\alpha}=(-1/2)g_{00},_{\beta \alpha}$,
$E _{0 \beta \alpha },_0=0$ and 
$E _{0 \beta 0 },_{\alpha}=(1/2)[ g_{00} (\ln \alpha),_{\beta}
]_{\alpha}$.
So
$$
(1/2)g^{mn} \left( g_{00},_{mn} - [g_{00}(\ln \alpha),_n],_m \right) =
  -4\pi \rho V_0 V_0 .
$$
For a weak field 
$$
 y^n_{\mu} y_n,_{\nu} \approx  \left\{
\begin{array}{l}
1 \mbox{ for} \mu = \nu =0 \\
-1 \mbox{ for} \mu = \nu \ne 0 \\
0 \mbox{  otherwise} ,
\end{array}
\right.
$$
and so
$$
g_{\mu \nu} = \alpha y^n_{\mu} y_n,_{\nu} = 
\pmatrix{
     \alpha & 0 & 0 & 0 \cr
     0 & -\alpha & 0 & 0 \cr
     0 & 0 & -\alpha & 0 \cr
     0 & 0 & 0 & -\alpha 
    }
, \; 
g^{\mu \nu} = 
\pmatrix{
     1/\alpha & 0 & 0 & 0 \cr
     0 & -1/\alpha & 0 & 0 \cr
     0 & 0 & -1/\alpha & 0 \cr
     0 & 0 & 0 & -1/\alpha 
    }.
$$
For a static and weak field $g^{00}V_0 V_0 = 1/\alpha$, so $V_0=1$ and 
\begin{equation}
g_{00},_{mm} -[g_{00}(\ln \alpha),_m]_m = 8\pi \rho .
\label{Einmod}
\end{equation}
From (\ref{phidefn}) the covariant potential $\phi $ is such that
$(1/2)[(\ln g_{00}),_n - (\ln \alpha),_n]= \phi ,_n$, and so
substituting this into (\ref{Einmod}) gives
\begin{equation}\label{eq:19}
(g_{00} \phi ,_m) ,_m = 4\pi \rho,
\label{newgrav}
\end{equation}
and rearranging (\ref{phidefn}) gives
$$
\frac{g_{00}}{\alpha} = 1 +2 \phi.
$$
It is seen that although $g_{00} /\alpha $ must be close
to unity for weak distortion, $g_{00}$ is unrestricted.
So $g_{00}$ can be equated to the MOND function, 
retrieving MOND dynamics in a simple and straightforward way.

For a point source, following the same arguments as \cite{Bekenstein:1984}, from (\ref{eq:19}) we get Newton's second law given explicitly as
\begin{equation}
M g_{00}{\bf a} = {\bf F}
\end{equation}
assuming no curl vector field present, where $M$ is the point mass, ${\bf a}$ the acceleration and ${\bf F}$ the force, which is the standard MOND modification but with $g_{00}$ identified as the MOND interpolation function $\mu$.

\section{Point source and general solution}

Using (\ref{phidefn}) and substituting $g_{00} = \alpha e^{2\phi}$ into (\ref{newgrav}) yields after integration that 
\begin{equation}\label{PS_1}
\alpha e^{2\phi} = \frac{D}{\left( r^2 | \tilde{\bigtriangledown} \phi|\right)}
\end{equation}
for point source of mass $M$,  $\rho = M \delta (r)$, where $D$ is an integration constant, $r= \sqrt{x_1^2+x_2^2+x_3^2}$ and $\delta (r)$ is the Dirac delta function. Matching this solution to the observed flattening of galaxys' rotation curves imposes that $| \tilde{\bigtriangledown} \phi| \rightarrow D/r^2$ when $| \tilde{\bigtriangledown} \phi| \gg a_0$  and that $| \tilde{\bigtriangledown} \phi| \rightarrow \sqrt{a_0 D}/r$ when $| \tilde{\bigtriangledown} \phi|  \ll a_0$, where $a_0$ is the acceleration parameter of MOND theory. Thus a consistent solution for the potential is derived to be
\begin{equation}
\phi = -M/r + \sqrt{a_0 M} \ln r,
\label{pointphi}
\end{equation}
where $D$ has been identified as the point source mass. This empirical derivation allows interpretation of the rate of expansion, suggesting a physical context and thus an alternative derivation (see later). The first term is the Newtonian potential due to the curvature $\phi ^{NEWT}$, and the second term is the MOND potential due to local expansions $\phi ^{MOND}$, see figure 1. Then, $(g_{00} \phi, _m),_m = 4 \pi \rho$ means that $g_{00} \phi,_m = (M/r^2) \vec{r}$. So

\begin{figure}
\begin{center}
\includegraphics[width=12cm]{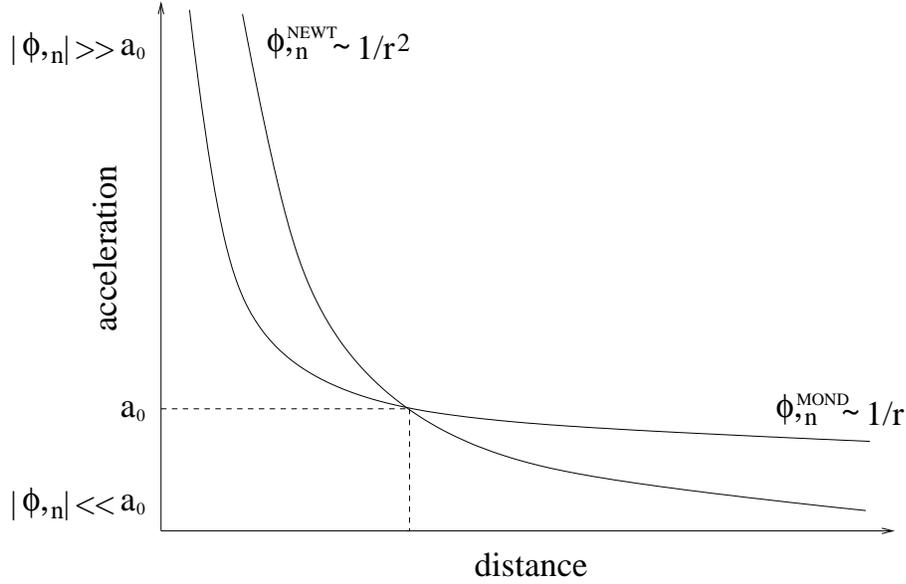}
\caption{The change in acceleration with distance}
\end{center}
\end{figure}

\begin{equation}
g_{00} = \frac{M/r^2}{M/r^2+\sqrt{a_0M}/r},
\label{pointg}
\end{equation}
and $g_{00}/\alpha = 1-2M/r+2\sqrt{a_oM}\ln r$. 
Two limits can now be considered.

For small $r$ such that the curvature term $M/r^2$ dominates the
expansion term $\sqrt{a_0M}/r$, 
then this equates to a dominant solution of the Newtonian potential
$\phi ^{NEWT}$ where the accelerations are such that $| \phi^N,_m|/a_0
>>1$.
Then (\ref{pointphi}) becomes
$\phi ^{NEWT} = -M/r$, $\phi ^{NEWT} ,_m = (M/r^2) \vec{r} $, 
and (\ref{pointg}) becomes
$g_{00} = 1-\sqrt{\frac{a_0}{M}} r \approx 1$,
$g_{00}/\alpha = 1-2M/r$. 
So, the Newtonian point source potential is recovered.

For large $r$ such that the expansion term $\sqrt{a_0M}/r$ dominates
the curvature term $M/r^2$,
then this equates to a dominant solution of the MOND potential
$\phi ^{MOND}$ where the accelerations are such that $| \phi^M,_m|/a_0
<<1$.
Then (\ref{pointphi}) becomes
$\phi ^{MOND} = \sqrt{a_0M}\ln r$, $\phi ^{MOND} ,_m = (\sqrt{a_0M}/r) \vec{r} $, 
and (\ref{pointg}) becomes
$g_{00} = \sqrt{\frac{M}{a_0}} \frac{1}{r} $.

Substituting into (\ref{newgrav}), gives
$[\sqrt{\frac{M}{a_0}} \frac{1}{r} (\sqrt{a_0 M} \ln r),_m ],_m =
[\frac{M}{r^2} \vec{r}],_m = 4\pi M \delta (x) = 4\pi \rho$ as
expected.

Also if limits are introduced directly into (\ref{PS_1}) such that for the Newtonian case as $r \rightarrow 0$, $\alpha = 1$ and $2\phi \ll 1$, this gives 
\begin{equation}
r^2 \left( 1+2\phi \right) \tilde{\bigtriangledown} \phi = M.
\end{equation}
After integration and $|\phi^2| \ll |\phi|$ yields $\phi = -M/r$ as expected.

In the MOND limit $|\phi^2| \gg |\phi^3|$ and $\alpha = \alpha(r)$. After integration this gives $\phi(r) \propto  \sqrt{a_0 M} \ln r$, where 
\begin{equation}
\alpha(r) = \frac{1}{2a_0r\ln r}.
\end{equation}
Interestingly, the $1/ r\ln r$ dependence for $\alpha$ (the space-time expansion) is identical to the large r radial velocity of a spherical shock wave \cite{Landau:1945} \cite{Whitham:1927} \cite{Smoller30092003}. So, if one assumed this physical origin for expansion one can directly derive the second term in (\ref{pointphi}) without fitting MOND characteristics to the solution. 

It is noted that the factor $g_{00}$ is approximately unity in the
Newtonian approximation, meaning that (\ref{newgrav}) is linear and so
a system of point sources can be considered as a summation of
separate point source solutions.
However, in the MOND approximation, $g_{00}$ is a varying function,
and so (\ref{newgrav}) is non-linear and cannot be broken down in this
way.
Furthermore, the mass term on the right hand side of (\ref{newgrav})
is split into a factor $\sqrt{M}$ with the potential and a factor
$\sqrt{M}$ with $g_{00}$.
So, the momentum equation of Newton's second law only makes sense if
it is modified to include the factor $g_{00}$.  
Furthermore, because of the nonlinearity this factor $g_{00}$ can only
be calculated once the complete system is known.

The point source solution suggests a general solution given by
\begin{eqnarray}
\phi &=& \phi ^{NEWT} + \phi ^{MOND} \nonumber \\
g_{00} &=& \left| \frac{\nabla \phi ^{NEWT} }{\nabla \phi ^{NEWT} + \nabla \phi
  ^{MOND} } \right| \nonumber \\
\alpha &=& \frac{1-2\phi ^{NEWT} -2 \phi ^{MOND}}{1+|\nabla \phi ^{MOND} /\nabla \phi
  ^{NEWT} |} \nonumber \\
g_{00}/\alpha &=& 1+2\phi ^{NEWT} +2\phi ^{MOND} \nonumber
\end{eqnarray}
where $\nabla $ is the differential operator $(
\frac{\partial }{\partial x_1}, 
\frac{\partial }{\partial x_2}, 
\frac{\partial }{\partial x_3})$
for Cartesian co-ordinate system vector representation $(x_1 ,x_2,
x_3)$. $\phi^{NEWT}$ and $\phi^{MOND}$ are connected in the sense that they can be seen as limiting values of the same general potential $\phi$, such that the first is the limit of small relative radius for solar systems, and the second is the limit of large relative radius for galaxies. So this choice of $\phi$ has a certain degree of physical justification in that it gives the expected physics in these limits.
The two limits are then as follows.

When $|\phi ,_n|/a_0 >> 1$, then curvature dominates so
$|\nabla \phi ^{NEWT} | >> |\nabla \phi ^{MOND} |$, 
and 
\begin{eqnarray}
\phi &=& \phi ^{NEWT} \nonumber \\
g_{00} &=& 1 \nonumber \\
\alpha &=& 1 - 2 \phi ^{NEWT} \nonumber \\
g_{00}/\alpha &=& 1+2 \phi ^{NEWT} , \nonumber
\end{eqnarray}
and so $\phi ^{NEWT} ,_{mm} = 4\pi \rho$, and the Newtonian gravitational
representation is recovered.
Such accelerations feature in solar system dynamics.

However, when $|\phi ,_n|/a_0  << 1$, then expansion dominates
$|\nabla \phi ^{MOND} | >> |\nabla \phi ^{NEWT}|$,  
and 
\begin{eqnarray}
\phi &=& \phi ^{MOND} \nonumber \\
g_{00} &=& \left| \frac{\nabla \phi ^{MOND}}{a_0}  \right| \nonumber \\
\alpha &=& \left| \frac{\nabla \phi ^{MOND}}{a_0} \right| (1-2\phi ^{MOND}) \nonumber \\
g_{00}/\alpha &=& 1+2 \phi ^{MOND} , \nonumber
\end{eqnarray}
and so (\ref{newgrav}) becomes
$$
(g_{00} \phi ,_m ),_m = ( \frac{|\nabla \phi ^{MOND} |}{a_0} \phi ^{MOND} ,_m),_m =
  4\pi \rho,
$$
which is the MOND representation for the potential acceleration.
Such accelerations feature in the motions of galaxies.

\vspace{9pt}

{\bf Acknowledgments}
This work was undertaken through an STFC studentship grant at the
University of Salford. We are grateful for the helpful comments of the referees. 

 
\bibliography{IEEEabrv,chadwick}

\end{document}